# Out-of-plane orientated self-trapped excitons enabled polarized light guiding in 2D perovskites

Junze Li[1], Junchao Hu[1], Ting Luo[1], Dongliang Chen[2], Yingying Chen[1], Zeyi Liu[1], Dingshan Gao[2], Xinglin Wen[1,2*] and Dehui Li[1,2*]

[1]*School of Optical and Electronic Information, Huazhong University of Science and Technology, Wuhan, 430074, China.*
[2]*Wuhan National Laboratory for Optoelectronics, Huazhong University of Science and Technology, Wuhan, 430074, China.*

*Correspondence to: wenxl@hust.edu.cn; dehuili@hust.edu.cn.

**Abstract**

Active optical waveguides combine light source and waveguides together in an individual component, which are essential for the integrated photonic chips. Although 1D luminescent materials based optical waveguides were extensively investigated, 2D waveguides allow photons to flow within a plane and serve as an ideal component for the ultracompact photonic circuits. Nevertheless, light guiding in 2D planar structures normally relies on the precise control of molecular orientation, which is complicated and low yield. Here, we report a strategy to guide polarized light in 2D microflakes by making use of the out-of-plane (OP) orientation of self-trapped excitons in as-synthesized 2D perovskite microplates. A space confined crystallization method is developed to synthesize 2D perovskite microflakes with dominated broad self-trapped excitons emission at room temperature, which are highly OP orientated with a percentage of the OP component over 85%. Taking advantages of the negligible absorption coefficient and improved coupling efficiency of OP orientated self-trapped exciton emission to the planar waveguide mode of the as-synthesized perovskite microflakes, we have achieved a broadband polarized light guiding with a full width at half maximum over 120 nm. Our findings provide a promising platform for the development of ultracompact photonic circuits.

**Introduction**

Luminescent material based active waveguides can produce gain or nonlinear optical effects for signal amplification, which are useful in optical information processing and communication interconnection.[1-3] More importantly, active optical waveguides also play a critical role for the development of integrated photonic circuits by combining light sources and waveguides together in an individual component.[4] Perovskites exhibit excellent optoelectronic and photonic properties simultaneously,[5-8] making them potential candidates to construct active waveguides. 1D perovskites wires have been demonstrated to guide light.[9-11] However, they were achieved by the self-absorption and re-emission process. 2D waveguides allow photons to transport within a plane, which can serve as an ideal component to control the photons flow for applications of ultracompact photonic chips. 2D waveguides based on planar perovskite plates have been demonstrated via edge emission , which was attributed to ballistic transport of exciton-polaritons under strong coupling condition[12] or whispering gallery modes (WGM).[13] However, the formation of whispering gallery modes requires a regular shape and light only circulates along surface. Furthermore, self-absorption/re-emission enabled waveguiding and ballistic transport of exciton-polaritons are usually isotropic within a planar photonic structure, making it challenge to control the direction of light flow. Currently, anisotropic light transport in 2D photonic structures can be realized by controlling the orientation of organic molecules,[14, 15] which brings the drawbacks of complicated synthesis and low yield. Therefore, it is desirable to exploit new mechanism to construct robust 2D active optical waveguide.

2D lead halide perovskites are a class of layered materials and exhibit unique optical properties due to their naturally formed quantum well structure and great tunability of their electronic band structure.[16, 17] Their bandgap can be easily tuned in the whole visible wavelength range by changing the layer number and compositions while a large exciton binding energy and an intense exciton emission can be achieved at room temperature due to the dielectric confinement effect.[18-22] Furthermore, the physical properties of 2D perovskites can be widely tuned via simply changing the species of long-chain spacers and inorganic layers.[20] More importantly, excitons in 2D perovskites has an out-of-plane (OP) component.[23-25] Theoretical calculations indicate that emissive dipoles with OP orientation can couple with planar waveguides with efficiency an order of magnitude larger than in-plane (IP) dipoles.[26, 27] Nevertheless, the OP component of the excitons in 2D perovskites in existed literatures is smaller than 18%, despite it can be tuned by changing the inorganic layer thickness and organic molecules.[12, 28] The soft nature of lattice together with the strong electron-phonon interaction leads to the formation of self-trapped excitons (STE) in 2D perovskites, which manifest them via large Stokes shift, broad emission band and negligible absorption.[29] In particular, free excitons (FE) dipoles exhibit dominated IP component with a small portion of OP one, while the broadband STE emission is closely related to the OP tilting of inorganic octahedral, which can be tuned by selecting the organic spacer in 2D perovskites.[30] Thus, it is expected that STE might exhibit OP dipole moment transition to efficiently couple to 2D planar waveguide; nevertheless, study on this aspect remains elusive.

Here, we developed a space confined crystallization method by using mica as templates to synthesize $(C_8H_9NH_3)_2PbI_4$ (abbreviated as $(PEA)_2PbI_4$) 2D perovskite microflakes with dominated broadband STE emission at room temperature. The STE is highly OP orientated with a percentage of the OP component over 85% measured by momentum-resolved PL. Taking advantages of negligible absorption coefficient and efficient coupling of OP orientated STE emission to waveguide mode of 2D perovskites, we demonstrate a broadband polarized light waveguide in 2D flakes with a bandwidth over 120 nm. We for the first time reveal the electric dipole of STE is along OP direction, which makes it feasible to couple with the TM waveguide mode to enable the polarized light guiding. As a result, we realized waveguide function and light emission simultaneously in an individual perovskite plate to achieve an active planar waveguide.

**Results**

**Synthesis and characterizations of 2D perovskite flakes**

A space confined crystallization method was developed to synthesize $(PEA)_2PbI_4$ 2D perovskite microflakes (Methods and Fig. 1a). Briefly, the solution synthesized $(PEA)_2PbI_4$ crystals were dissolved into a mixed solution of hydroiodic acid and hypophosphorous acid, which then was dropped onto a hot glass substrate and a mica flake (top cover) was pressed onto the droplet under continuously heating at 80 ºC for crystallization. For convenience, Mica-P is used to stand for the samples grown by using mica flakes as covers hereafter. Powder X-ray diffraction (XRD) pattern for Mica-P is consistent with the theoretical calculated one and can be indexed to (00*l*) direction (Fig. 1b). Fig. 1c displays the optical and fluorescence images of Mica-P, which shows a smooth surface and uniform red fluorescence. It is important to be noted that the relative concentration of HI significantly affects the emission properties of the as-synthesized samples (Fig. S1).

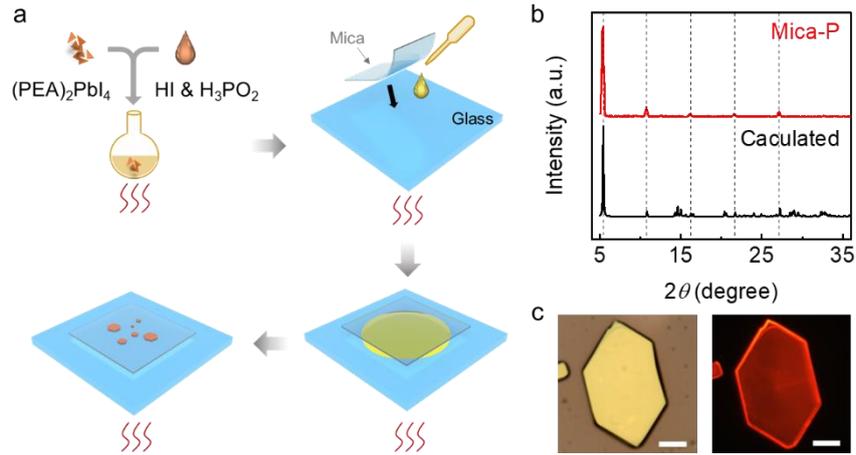

***Fig.1 Characterizations of synthesized perovskites microflakes.*** *(a) Schematic of the space confined crystallization method. (b) Powder XRD patterns of Mica-P and calculated one. (c) Optical (left) and fluorescence (right) images for Mica-P. Scale bars are 20 μm.*

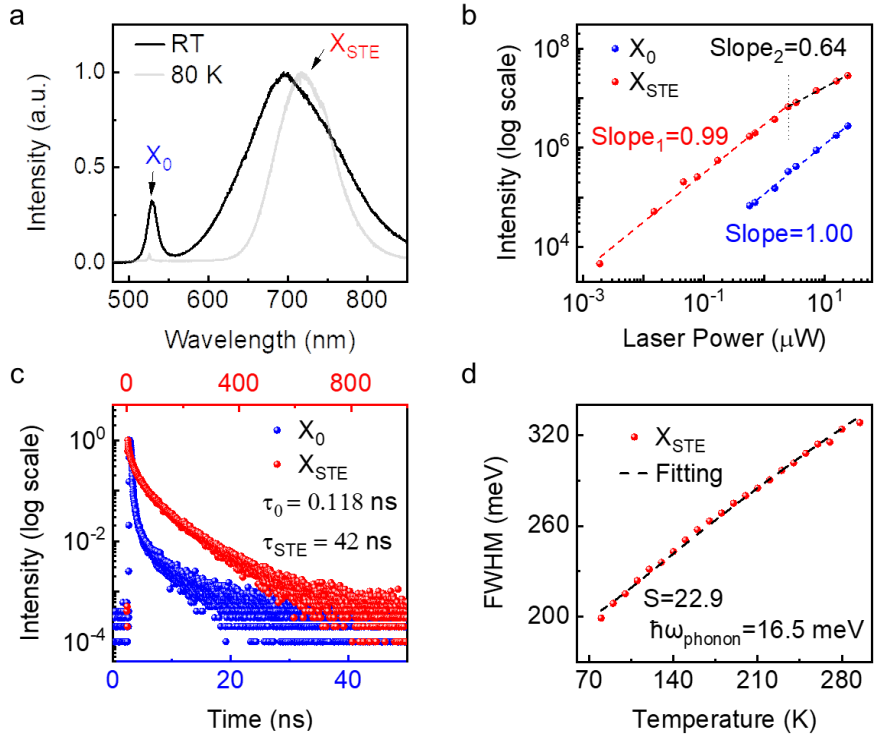

***Fig.2 Characterizations of self-trapped excitons.*** *(a) PL spectra at room temperature and 80 K. Narrowband and broadband peaks are marked with $X_0$ and $X_{STE}$, respectively. (b) PL intensity of $X_0$ and $X_{STE}$ versus excitation power. The dashed lines are the result of linear fits. (c) Time-resolved PL decays for $X_0$ and $X_{STE}$. (d) The FWHM of $X_{STE}$ as a function of temperature. The dashed line is the fitting result.*

The photoluminescence (PL) spectra of Mica-P at room temperature show that below the free exciton ($X_0$) emission peak, there is an additional broadband emission peak (denoted as $X_{STE}$) (Fig.

2a). The PL results were consistent with the red color of fluorescence images in Fig. 1c. At 80 K, $X_{STE}$ is enhanced compared with $X_0$. According to previous reports together with the large Stokes shift and broad emission band, $X_{STE}$ can be assigned to the STE emission.[31, 32] As the excitation power increases, the emission intensity of $X_0$ and $X_{STE}$ linearly increases with excitation power before saturation (Fig. 2b and Fig. S2) and the emission energy remains the same, supporting that $X_{STE}$ emission peak is from STE emission rather than doping and/or defects.[33, 34] Furthermore, $X_0$ exhibits a short lifetime ~0.118 ns while $X_{STE}$ shows a much longer lifetime ~42 ns (Fig. 2c and Fig. S3), which is comparable to that of reported STE in 2D perovskites and is another typical feature of STE in 2D perovskites.[35-37] In addition, the absorption spectra show that only photons with energy above $X_0$ can be absorbed, which provides another strong evidence for the assignment of STE again (Fig. S4).[33] In particular, the estimated Huang-Rhys factors, characterizing the strength of electron-phonon coupling via temperature-dependent full width at half maximum (FWHM) of PL (Fig. 2d, Fig. S5 and Note S1), is 22.9, which is much larger than the critical value allowing STE to be formed in 2D perovskites.[33] Therefore, all those evidences point to that the broad emission peak is originated from STE emission.

**Dipole orientation of STE**

Considering that STE is closely related to the OP tilting of inorganic octahedra,[30] we expect a large proportion of OP oriented component for STE. The back focal plane imaging system was adopted to determine the orientation of luminescent excitons in layered materials (Fig. S6).[38-40] We first verified the accuracy of the measurement system by using monolayer $WS_2$, whose FE is completely IP orientated, agreeing excellently with previous reports (Fig. S7).[41] We then measured the orientations of both FE and STE emission of 2D perovskites by utilizing a narrow bandpass filter to select the peak band in order to reduce the dispersion effect from optical elements (Fig. S8).

Fig. 3a displays the FE emission patterns of Mica-P in momentum space, where $k_x$ and $k_y$ correspond to the horizontal and vertical components of the light wavevector, respectively. The intensity in Fig. 3a and Fig. 3b are normalized and they share same scale bar so that it is easier to carry out data fittings. The absolute intensity of STE is much higher than FE as shown in Fig. 2a. The stronger the intensity of p-polarized component at $|k_{//}/k_0|=1$, the larger the proportion of OP excitons.[39] For FE, the p-polarized component near $|k_{//}/k_0|=1$ is not completely vanished, which means that FE dipole moment contains both IP and OP components but dominates by IP one. In contrast, as can be seen from Fig. 3b, the p-polarized component of STE exhibits a strong luminescence near $|k_{//}/k_0|=1$, indicating a large OP exciton fraction. We fitted the s- and p-polarized emission based on the model that is well established in previous reports (Note S2).[38-40]

A simultaneous fit of s- and p-polarized component reveal a 70% IP excitons emission for FE (Fig. 3c and Fig. S9c), while the fitting shows that STE only have 15% IP exciton component, which is much smaller than that of FE (Fig. 3d and Fig. S9d). It is import to note that curves in Fig. 3c and Fig. 3d have rather different features. We put the p-polarized emission in Fig. 3c (FE) and Fig. 3d (STE) together for comparison and the difference can be clear visualized at $|k_{//}/k_0|=1$ (Fig. S10). It is worth noting that the experimental data agree with the simulated momentum space emission patterns in the range $|k_{//}/k_0|\leq 1$. However, the fitting errors are large in the momentum space range $|k_{//}/k_0|>1$. This can be attributed to the effects of imaging artifacts observed at higher numerical aperture values together with the lower effective numerical aperture of the coupled microscope-lens-spectrograph system.[39] The slight difference of radius of FE (Fig. 3a) and STE (Fig. 3b) is attributed to the deviated focus condition (Fig. S11). In our measurement, the weak free exciton emission makes it difficult to focus accurately, leading to a slight radius deviation. The slight difference of radius of the ring has minimum effect on our interpretation.

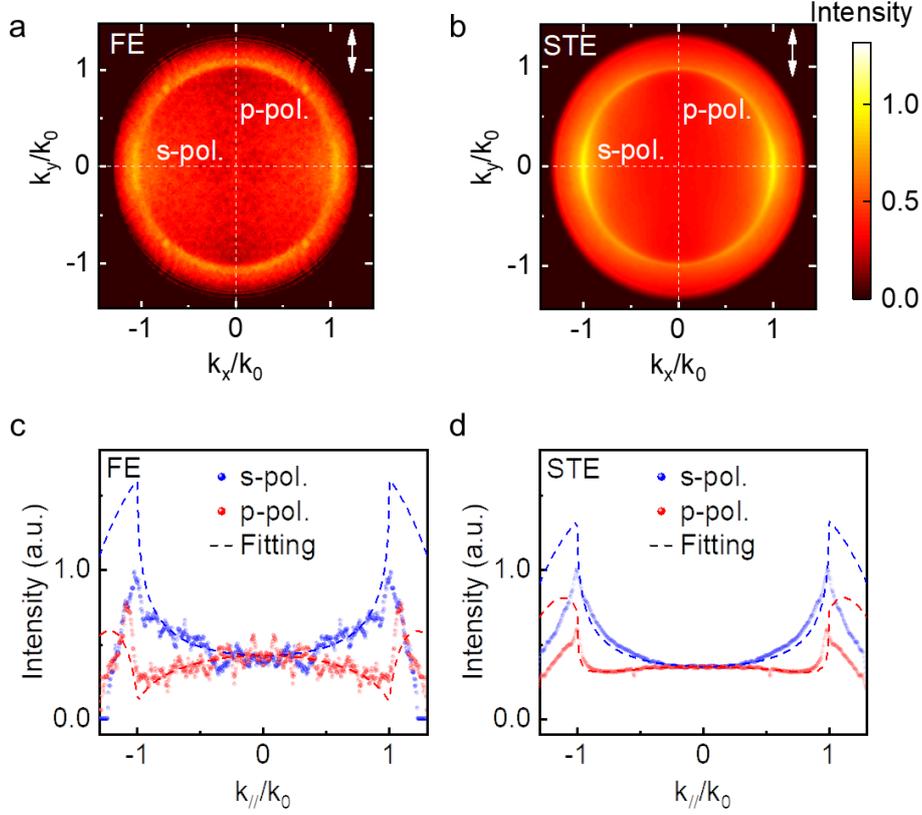

***Fig.3 Luminescent orientation of STE.*** *(a, b) Normalized momentum space emission pattern of FE (a) and STE (b). The white arrow in the upper right corner is the direction of the transmission axis of the polarizer. The s- and p-polarized cross-sections are marked by white dashed lines. (c, d) Experimental s- (blue) and p- (red) polarized cross-sections of FE (c) and STE (d) momentum spectra emission pattern. The blue and red dashed lines represent fittings of the experimental data.*

The IP component of FE accounts for about 70%, which is comparable to that in previously reported works.[12] Strikingly, STE is highly OP oriented with a percentage over 85%, which is much larger than that in 2D layered materials [28, 39, 41] except InSe.[42] Given that the twisting of the inorganic octahedra in the OP direction promotes the generation of STE and the STE is closely related to organic chain spacer,[43] we speculate that the OP oriented dipole moment of the STE is driven by the organic chain twisting along the OP direction. Nevertheless, further investigations and simulations are required to fully clarify the underlying mechanism. It should be noted that the highly OP oriented

STE is rather different from OP oriented fine structure of band-edge excitons previously reported in (PEA)$_2$PbI$_4$ 2D perovskites.[24, 44] All OP and IP fine structure splitting excitons are categorized as free excitons, which correspond to the free exciton emission (X$^0$, ~ 528 nm) we observed in Fig. 2a. Because the energy splitting is as small as 1.3 meV,[24] the IP and OP free excitons could not be distinguished in our measurement at 77 K.

**Polarized light guiding**

Guiding light with specific polarization is important for polarization multiplexing technique and optical communication interconnection. Though polarized light guiding has been demonstrated in 1D wires,[2, 45] it is of great importance to control the flow of polarized light in a 2D plane for ultracompact photonic circuits. Taking advantages of the negligible absorption of STE and OP orientation of STE that could efficiently couple emission light into waveguide modes of 2D perovskite microplate, we have achieved robust polarized light guiding by using 2D perovskite microplates.

Emission along the edges of bare perovskite flakes is not homogeneous because of the deviated scattering efficiency at different position on the edge (Fig. S12). Therefore, a ring-shaped structure on a 300 nm SiO$_2$/Si substrate has been specially fabricated with as-synthesized 2D perovskite microflake covered on it to obtain a homogeneous edge emission for better visuality of polarized light guiding (Fig. 4a). A thin layer of polymethylmethacrylate (PMMA) was spin-coated between the Au ring and 2D perovskite microflake to block charge transfer while a layer of boron nitride (h-BN) was transferred on top of the device to protect the 2D perovskite microflake from being degraded during the measurement. The ring-shape structure was used to outcouple the propagating light within microflake via bending the microflake so that the emitted light from OP oriented excitons can be detected.

PL images of FE and STE are displayed in Fig. 4b and Fig. 4c when the microflake was excited at the center of the ring. The PL images show that FE emission can be only observed at the excitation point, whereas STE emission is present both at the excitation point and at the ring. The absence of FE emission at the ring can be ascribed to the conversion of FE to STE and low coupling efficiency of IP dipole emission to microplate, while the propagation of STE emission toward the ring is due to efficient coupling between emitted light of OP oriented excitons and the planar waveguide mode of microflake (Fig. S13). Compared with IP oriented exciton emission, the coupling efficiency for OP oriented excitons can be improved more than 190% (Fig. S14), exhibiting great potential for on-chip integrated planar integrated photonics. Simulations indicate that perovskites microplate sustains $TM_0$ and $TE_0$ waveguide modes. Electric field of $TM_0$ mode is confined inside perovskite plate (Fig. S15b), while most of the electric field of $TE_0$ mode dissipate to PMMA layer (Fig. S15c). Therefore, we believe that the STE emission predominantly couples to $TM_0$ waveguide mode. In addition, the polarization of radiation of OP electric dipole is always parallel to the propagation plane (Fig. S16). As a result, the OP oriented STE emission can couple to $TM_0$ waveguide mode efficiently.

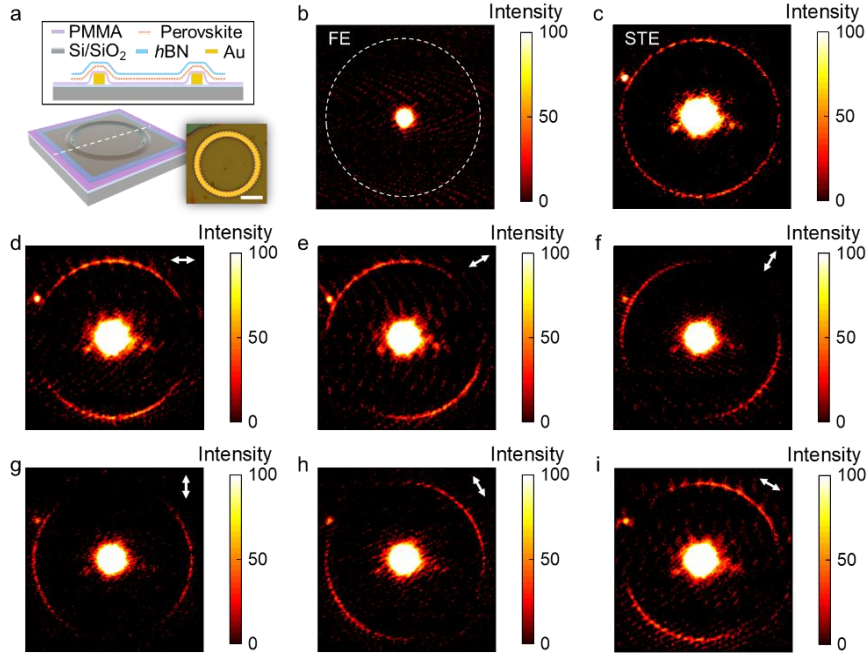

***Fig.4 Polarized light guiding.*** *(a) Schematic diagram of the device for guiding polarized light. The top is the cross-sectional view of the dashed-line marked area and inset in the lower right corner is optical image. Scale bar is 10 μm. (b, c) PL images for FE (b) and STE (c). FE and STE images were measured with a 600 nm short pass filter and a 600 nm long pass filter, respectively. (d-i) Polarization resolved PL images of STE. The white arrow is the direction of the transmission axis of the polarizer. The intensity of the excitation point at the center of the ring is saturated.*

We have further acquired the PL images under different polarization direction (Fig. 4d-i). It was observed that the pattern of the ring-shaped luminescence rotates when rotating the transmission axis of polarizer (Fig. 4d-i). STE emission propagate along the waveguide in TM mode (Fig. S15b), while we observed a TE-like far field that polarization is always perpendicular to the light propagation plane. It might be the scattering at the position of Au ring that causes a polarization change.[46-48] Furthermore, emission intensity as a function of the angle between the propagation path and the transmission axis of the polarizer was measured (Fig. 5). When the propagation path is perpendicular to the transmission axis of polarizer, the intensity reaches its maximum, and luminescence vanishes when they are parallel. The polarization dependent emission feature is further verified by the PL spectrum at the ring, which also shows a broadband emission with FWHM of 120 nm (Fig. S17). Those observations suggest that the emitted polarized light propagate at a specific route. These results clearly indicate that the polarization direction of the broadband luminescence with a bandwidth of 120 nm is always perpendicular to its propagation path in planar and thus polarized light guiding is achieved.

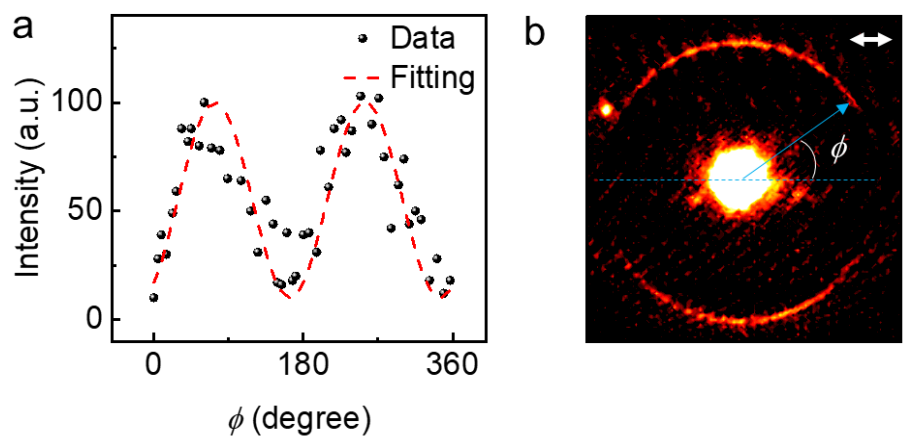

*Fig.5 Angle dependent intensity.* (a) The intensity along the ring extracted from (b) as a function of the angle ($\phi$) between the propagation path and the transmission axis of the polarizer. The red dotted line is the fitting via the formula: $I = I_0 + A\cos(\phi - \theta)$, where $I_0$ is 55, $A$ is 45 and $\theta$ is 28.

It should be noted that the ring-shaped structure is not necessary to observe the polarized light guiding. Similar broadband polarized light emission has also observed at the edge of the synthesized Mica-P far away from the excitation point (Fig. S18), suggesting that the polarized light guiding is intrinsic feature of the propagating light within the perovskite flakes. In addition, by properly selecting the chemical compositions and layered number of 2D perovskites, the emission energy of STE in 2D perovskites can be tuned in the whole visible wavelength range and thus the broadband polarized light guiding can be realized in the whole visible wavelength range.[30, 49] Therefore, our study provides an efficient and simple route to achieve on-chip broadband polarized light guiding in 2D planar structure for photonic applications.

**Discussion**

Fieramosca et al also observed edge emission of (PEA)$_2$PbI$_4$ 2D perovskite microflakes.[12] The key difference between our work and Fieramosca's is that our work investigates the STEs in weak coupling regime while Fieramosca et al studied exciton polaritons (EPs) formed under strong coupling condition. Therefore, the emission spectrum of our sample is distinctively different from their work. Whereas in their sample the emission spectrum is dominated by free exciton emission (2.35 eV, corresponding to 528 nm), the broad emission in our sample is mainly from STE emission, locating at 700 nm with a large Stokes shift of ~170 nm (Fig. S4). In addition, the emission from the edges in Fieramosca's work is ascribed to the ballistic transport of exciton polaritons. In our case, STE emission peak is much far below the absorption band edge together with out-of-plane orientated dipole moment of STEs, the emitted light from STEs can efficiently couple to the waveguide mode of perovskite plate without self-absorption. STEs cannot absorb light because self-trapped states can be regarded as excited states[29], these excited states disappear once the excitation is removed (Fig. S4, absorption spectrum). As a result, the STE emission can propagate toward the edge by coupling to waveguide modes of our sample. To sum up, in Fieramosca's work the exciton polaritons propagate to the edge and then emit light there, while in our case the STE emission takes place at the excited point and the emitted light propagate to the edge to be observed. Therefore, the exciton species and emission mechanism are both different between our case and Fieramosca's work.

In addition, our work is different from the edge emission in WGM resonators. Energy is confined around the surface for WGM resonators, namely, light travels along the surface.[6, 50, 51] Energy leakage from the surface leads to the observation of edge emission. Self-absorption and re-emission involve when light travel in WGM resonators. Waveguide mode in our work is ascribed to the total internal reflection at the top and bottom surface rather than the lateral surface, which makes that a regular shape is not necessary. More importantly, the light emission in our sample takes place at the excited point and then propagate to the edge of sample owing to the free absorption for the STE emitted light.

It is also important to exclude the EPs ballistic transport or exciton diffusion here. FWHM of STEs in our work (Fig. 2d) is larger than the coupling strength (170 meV) of $(PEA)_2PbI_4$ reported by Fieramosca[12], indicating the strong coupling between STE and photons is not satisfied. In addition, STEs are excited states and thus cannot absorb light at all.[29] Therefore, exciton-polaritons transport can be excluded. It has been reported that diffusion length of free exciton is about $236 \pm 4$ nm for $(PEA)_2PbI_4$.[52] STEs should exhibit shorter diffusion length than free exciton due to the localization induced by strong exciton-phonon coupling.[29] In contrast, the distance between the excited point and the edge of sample can be as large as tens of micrometers in our case, which thus can exclude the possibility of the diffusive transport of excitons. Additionally, there should have fluorescence signal along the diffusion path in the case of exciton diffusion, while we only observed emission at the edge in Fig. 4. All those evidences support that the EPs ballistic transport or exciton diffusion can be ruled out in our case.

In summary, we have synthesized 2D perovskite microflakes with strong STE emission by a space confined crystallization method. Using the back focal plane imaging, we found that the OP component of STE accounts for 85%, which is much larger than 30% of free excitons. The highly OP orientation of STE might be ascribed to the tilting of the inorganic octahedra in the OP direction caused by the twisting of the organic chains. Benefiting from the large proportion of OP oriented excitons and negligible absorption coefficient of STE, broadband emission can couple to the microflake waveguide more efficiently, leading STE emission propagate along the microflake. Furthermore, by stressing the perovskite with a ring substrate, we realized a polarized light guiding in a 2D planar structure. Our study shows that 2D perovskites can provide a platform for the study of light-matter interactions where OP exciton is critical and would also shed light on ultracompact integrated photonic chips.

## Materials and methods

**Perovskite flake preparation.**

All chemicals were purchased from Sigma-Aldrich. (PEA)$_2$PbI$_4$ 2D perovskite crystals were synthesized by a solution method according to reported works.[30, 53] The precursor solution was obtained by dissolving (PEA)$_2$PbI$_4$ crystals (0.3 g) in a mixing solution contains 57% w/w aqueous hydroiodic acid (HI) (4 mL) and 50% w/w aqueous hypophosphorous acid (H$_3$PO$_2$) solution (1 mL) under stirring at 140°C for 1 hour. The pre-cleaned and dried glass substrates were heated to 80 °C on a hot plate. Then, precursor solution was dropped onto the glass substrate and a top cover (Mica) was pressed onto the droplet with heating continuously at 80 °C for 1 hour for the perovskite flakes to crystallize.

**Device fabrication.**

Au ring with a thickness of 50 nm was defined by electron beam lithography and deposited by thermal evaporation onto SiO$_2$/Si wafer. A thin layer of PMMA was spin-coated between the Au ring and 2D perovskite microflake to block charge transfer. Perovskite microflake was mechanically exfoliated from crystal with thickness around 100 nm. A thin boron nitride flake was then transferred on top of the device to protect the 2D perovskite microflake from being degraded during the measurement.

**Material characterizations.**

PL measurements were performed on a homebuilt Raman spectrometer (iHR-550 Horiba) with a 600 g/mm grating, whereas a 473 nm solid-state laser was used as the excitation source. The absorption measurements were carried out on the same setup using a halogen lamp as the light source. Time-resolved PL decays were carried out on Bentham MSH300 with a 1200 g/mm grating.

Thickness of the flakes was measured by stylus profilometry (Bruker DektakXT) and atomic force microscope (SHIMADZU, SPM9700).

**Back-focal-imaging spectroscopy and PL measurements.**

The schematic illustration of the measure setup for back-focal-imaging is shown in Supplementary Information Fig. S6. A set of lenses is used to send the back focal plane image of the 100× objective (Olympus, CFI Plan Fluor Oil, NA = 1.3) to the charge-coupled device (Q-imaging, RETIGA 20000R). A 473 nm/556 nm solid-state laser is used as excitation source and a long pass filter of 500 nm/600 nm (Thorlabs FEL0500 and FELH0600) is used to block the excitation light into the spectrometer. The desired wavelength signal is obtained by using narrow bandpass filters (Thorlabs FBH520-40 and FBH700-10). The perovskite flake is placed in air, and the back of the quartz substrate is immersed in oil. To obtain a reliable results, the polarizer is rotated 360° and momentum space images are acquired every 90°. The four images obtained are rotated and their intensity-weighted average was calculated. This process is used to mean out the imperfections and defaults present on the optics and reduce the effects of any possibly existing anisotropy of our set-up.[42]

A flip mirror can be lifted to switch from the back-focal plane imaging to PL mode. The main path is the same as the path for back-focal-imaging. The desired emission signal at the excitation point (or at the ring) is selected by a slit (Daheng Optics GCM56) in front of spectrometer. The PL spectrum is collected by a 100× objective (Olympus, MPlan N, NA = 0.9) and the spectrometer is an iHR320 (Horiba, Jobin Yvon) equipped with a charge-coupled device (Horiba, Syncerity).

**Acknowledgements**

D. Li. acknowledges the support from NSFC (62074064), the National Key Research and

Development Program of China (2018YFA0704403), and the Innovation Fund of WNLO. X. Wen acknowledges the support from NSFC (62005091). J. Li acknowledges the support from NSFC (62104074), the project funded by China Postdoctoral Science Foundation (2020M682399) and Postdoctoral Innovation Research Funding of Hubei Province. We thank the Testing Center of Huazhong University of Science and technology and the Center of Micro-Fabrication of WNLO for the support in device fabrication and AFM measurements. We thank the Testing Center of Huazhong University of Science and technology and the Center of Micro-Fabrication of WNLO for the support in device fabrication and AFM measurements.

## Author contributions

D. Li and X. Wen conceived the idea and guided the project. J. Li performed most of the experiments. J. Hu and T. Luo contributed to the synthesis of perovskites. Y. Chen and Z. Liu assisted in devices fabrications. J. Li built the experimental set-up with assistance from X. Wen. D. Chen and D. Gao helped in waveguide simulations. J. Li, X. Wen and D. Li wrote the paper. All authors discussed the results and commented on the manuscript.

## Conflict of interest

The authors declare no competing financial interest.

**References:**

1.  Zhou, B.;   Xiao, G. W.; Yan, D. P., Boosting Wide-Range Tunable Long-Afterglow in 1D Metal-Organic Halide Micro/Nanocrystals for Space/Time-Resolved Information Photonics. *Adv. Mater.* **33**, 2007571 (2021).

2.  Wang, X., et al., Ligand-protected metal nanoclusters as low-loss, highly polarized emitters for optical waveguides. *Science* eadh2365 (2023).

3.  Subbaraman, H., et al., Recent advances in silicon-based passive and active optical interconnects. *Opt. Express* **23**, 2487-2510 (2015).

4.  Xiang, C.;   Jin, W.; Bowers, J. E., Silicon nitride passive and active photonic integrated circuits: trends and prospects. *Photonics Res.* **10**, A82-A96 (2022).

5.  Sutherland, B. R.; Sargent, E. H., Perovskite photonic sources. *Nat. Photonics* **10**, 295-302 (2016).

6.  Fu, Y. P., et al., Metal halide perovskite nanostructures for optoelectronic applications and the study of physical properties. *Nat. Rev. Mater.* **4**, 169-188 (2019).

7.  Su, R., et al., Room temperature long-range coherent exciton polariton condensate flow in lead halide perovskites. *Sci. Adv.* **4**, eaau0244 (2018).

8.  Qi, X., et al., Photonics and Optoelectronics of 2D Metal-Halide Perovskites. *Small* **14**, 800682 (2018).

9.  Wang, Z. Y., et al., Wavelength-tunable waveguides based on polycrystalline organic-inorganic perovskite microwires. *Nanoscale* **8**, 6258-6264 (2016).

10. Mao, W. X., et al., Controlled Growth of Monocrystalline Organo-Lead Halide Perovskite and Its Application in Photonic Devices. *Angew. Chem. Int. Edit.* **56**, 12486-12491 (2017).

11. Dursun, I., et al., Efficient Photon Recycling and Radiation Trapping in Cesium Lead Halide Perovskite Waveguides. *ACS Energy Lett.* **3**, 1492-1498 (2018).

12. Fieramosca, A., et al., Tunable out-of-plane excitons in 2D single-crystal perovskites. *ACS Photonics* **5**, 4179-4185 (2018).

13. Zhang, Q., et al., High-Quality Whispering-Gallery-Mode Lasing from Cesium Lead Halide Perovskite Nanoplatelets. *Adv. Funct. Mater.* **26**, 6238-6245 (2016).

14. Liu, Y., et al., Orientation-Controlled 2D Anisotropic and Isotropic Photon Transport in Co-crystal Polymorph Microplates. *Angew. Chem. Int. Edit.* **59**, 4456-4463 (2020).

15. Yao, W., et al., Controlling the Structures and Photonic Properties of Organic Nanomaterials by Molecular Design. *Angew. Chem. Int. Edit.* **52**, 8713-8717 (2013).

16. Blancon, J. C., et al., Semiconductor physics of organic-inorganic 2D halide perovskites. *Nat. Nanotechnol.* **15**, 969-985 (2020).

17. Su, R., et al., Perovskite semiconductors for room-temperature exciton-polaritonics. *Nat. Mater.* **20**, 1315-1324 (2021).

18. Dou, L., et al., Atomically thin two-dimensional organic-inorganic hybrid perovskites. *Science* **349**, 1518-1521 (2015).

19. Grancini, G.; Nazeeruddin, M. K., Dimensional tailoring of hybrid perovskites for photovoltaics.

*Nat. Rev. Mater.* **4**, 4-22 (2018).

20. Smith, M. D.;  Connor, B. A.; Karunadasa, H. I., Tuning the luminescence of layered halide perovskites. *Chem. Rev.* **119**, 3104-3139 (2019).
21. Do, T. T. H., et al., Bright Exciton Fine-Structure in Two-Dimensional Lead Halide Perovskites. *Nano Lett.* **20**, 5141-5148 (2020).
22. Do, T. T. H., et al., Direct and indirect exciton transitions in two-dimensional lead halide perovskite semiconductors. *J. Chem. Phys.* **153**, 064705 (2020).
23. DeCrescent, R. A., et al., Bright magnetic dipole radiation from two-dimensional lead-halide perovskites. *Sci. Adv.* **6**, eaay4900 (2020).
24. Posmyk, K., et al., Exciton Fine Structure in 2D Perovskites: The Out-of-Plane Excitonic State. *Adv. Opt. Mater.* 2300877 (2023).
25. Palummo, M.;  Postorino, S.;  Borghesi, C.; Giorgi, G., Strong out-of-plane excitons in 2D hybrid halide double perovskites. *Appl. Phys. Lett.* **119**, 051103 (2021).
26. Davanço, M.; Srinivasan, K., Efficient spectroscopy of single embedded emitters using optical fiber taper waveguides. *Opt. Express* **17**, 10542-10563 (2009).
27. Jun, Y. C.;  Briggs, R. M.;  Atwater, H. A.; Brongersma, M. L., Broadband enhancement of light emission in silicon slot waveguides. *Opt. Express* **17**, 7479-7490 (2009).
28. Cinquino, M., et al., Managing Growth and Dimensionality of Quasi 2D Perovskite Single-Crystalline Flakes for Tunable Excitons Orientation. *Adv. Mater.* **33**, 2102326 (2021).
29. Smith, M. D.; Karunadasa, H. I., White-Light Emission from Layered Halide Perovskites. *Acc. Chem. Res.* **51**, 619-627 (2018).
30. Du, K. Z., et al., Two-dimensional lead(II) halide-based hybrid perovskites templated by acene alkylamines: crystal structures, optical properties, and piezoelectricity. *Inorg. Chem.* **56**, 9291-9302 (2017).
31. Wu, X., et al., Trap states in lead iodide perovskites. *J. Am. Chem. Soc.* **137**, 2089-2096 (2015).
32. Mauck, C. M.; Tisdale, W. A., Excitons in 2D organic-inorganic halide perovskites. *Trends Chem.* **1**, 380-393 (2019).
33. Luo, J., et al., Efficient and stable emission of warm-white light from lead-free halide double perovskites. *Nature* **563**, 541-545 (2018).
34. Yangui, A., et al., Optical investigation of broadband white-light emission in self-assembled organic-inorganic perovskite $(C_6H_{11}NH_3)_2PbBr_4$. *J. Phy. Chem. C* **119**, 23638-23647 (2015).
35. Cortecchia, D., et al., Polaron self-localization in white-light emitting hybrid perovskites. *J. Mater. Chem. C* **5**, 2771-2780 (2017).
36. Yu, J., et al., Broadband extrinsic self-trapped exciton emission in Sn-doped 2D lead-halide perovskites. *Adv. Mater.* **31**, e1806385 (2019).
37. Straus, D. B., et al., Direct observation of electron-phonon coupling and slow vibrational relaxation in organic-inorganic hybrid perovskites. *J. Am. Chem. Soc.* **138**, 13798-13801 (2016).
38. Benisty, H.;  Stanley, R.; Mayer, M., Method of source terms for dipole emission modification in

modes of arbitrary planar structures. *J. Opt. Soc. Am. A* **15**, 1192-1201 (1998).

39. Schuller, J. A., et al., Orientation of luminescent excitons in layered nanomaterials. *Nat. Nanotechnol.* **8**, 271-6 (2013).
40. Brotons-Gisbert, M., et al., Engineering light emission of two-dimensional materials in both the weak and strong coupling regimes. *Nanophotonics* **7**, 253-267 (2018).
41. Wu, K., et al., Identification of twist-angle-dependent excitons in $WS_2/WSe_2$ heterobilayers. *Natl. Sci. Rev.* **9**, nwab135 (2022).
42. Brotons-Gisbert, M., et al., Out-of-plane orientation of luminescent excitons in two-dimensional indium selenide. *Nat. Commun.* **10**, 3913 (2019).
43. Koegel, A. A., et al., Correlating broadband photoluminescence with structural dynamics in layered hybrid halide perovskites. *J. Am. Chem. Soc.* **144**, 1313-1322 (2022).
44. Dyksik, M., et al., Brightening of dark excitons in 2D perovskites. *Sci. Adv.* **7**, eabk0904 (2021).
45. Hu, W. L., et al., Optical Waveguide Based on a Polarized Polydiacetylene Microtube. *Adv. Mater.* **26**, 3136-3141 (2014).
46. Yang, J. J.; Hugonin, J. P.; Lalanne, P., Near-to-Far Field Transformations for Radiative and Guided Waves. *ACS Photonics* **3**, 395-402 (2016).
47. Naruse, M., et al., Polarization in optical near and far fields and its relation to shape and layout of white. *J. Appl. Phys.* **103**, 113525 (2008).
48. Demarest, K.; Huang, Z.; Plumb, R. J. I. t. o. A., An FDTD near-to far-zone transformation for scatterers buried in stratified grounds. *IEEE trans. Antennas Propag.* **44**, 1150-1157 (1996).
49. Cai, P.; Huang, Y.; Seo, H. J., Anti-Stokes ultraviolet luminescence and exciton detrapping in the two-dimensional perovskite $(C_6H_5C_2H_4NH_3)_2PbCl_4$. *J. Phy. Chem. Lett.* **10**, 4095-4102 (2019).
50. Dong, H. Y., et al., Materials chemistry and engineering in metal halide perovskite lasers. *Chem. Soc. Rev.* **49**, 951-982 (2020).
51. Tian, X. Y., et al., Whispering Gallery Mode Lasing from Perovskite Polygonal Microcavities via Femtosecond Laser Direct Writing. *ACS Appl. Mater. Inter.* **13**, 16952-16958 (2021).
52. Seitz, M., et al., Exciton diffusion in two-dimensional metal-halide perovskites. *Nat. Commun.* **11**, 2035 (2020).
53. Zhang, Q., et al., Controlled aqueous synthesis of 2D hybrid perovskites with bright room-temperature long-lived luminescence. *J. Phy. Chem. Lett.* **10**, 2869-2873 (2019).